\documentstyle[12pt,thmsa,sw20lart]{article}

\input tcilatex

\begin{document}

\author{N.Mebarki, A.Taleb, H.Aissaoui, N.Belaloui, M.Haouchine. \\
D\'epartement de Physique Th\'eorique,\\
Institut de Physique, Constantine, ALGERIA.\thanks{%
Permanent adress.}}
\title{Successive Toroidal Compactifications Of A Closed Bosonic Strings.\thanks{%
This work is supported by the Algerian Ministry of education and research
under contract D2501/01/17/93.}}
\maketitle

\begin{abstract}
Successive toro\"\i dal compactifications of a closed bosonic string are
studied and some Lie groups solutions are derived.\newpage\ 
\end{abstract}

\section{Introduction}

Our present understanding of the observed fundamental interactions is
encompassed, on the one hand, for the strong, weak and electromagnetic
interactions by the standard model and on the other hand for the
gravitational interaction by Einstein's classical theory of general
relativity which, however, can not be consistently quantized.

Although the success of some of the unified gauge theories (based on the
point-like quantum fields concept), there are too many arbitrary parameters
and some of the outstanding problems like the Higgs, spontanious symmetry
breaking mechanism, Kobayachi Moskawa metrix etc... are still unsolved.

The discovery in the summer of $1984$ by Green and Schwarz \cite{1} of the
unique anomaly free open superstring has once again spurred an enormous
interest in string theories as candidates for unified quantum theories of
all interactions and matter.

As opposed to point-like particles in ordinary field theories, the
fundamental constituents of string theories are $1$-dimensional objects. A
single classical relativistic string can vibrate in an infinite set of
normal modes, which, when quantized correspond to an infinite set of states
with arbitrary high masses and spins.

These theories can be consistently quantized for one specific dimension of
space-time only. This critical dimension is $26$ for the bosonic string
(open or closed) and $10$ for the superstring \cite{2}, \cite{3}. However,
to keep contact with the real world, the extra space-time dimensions have to
be compactified. It turns out that there are too many ways to do a such
procedure and consequently, the four-dimensional low energy physics is not
unique \cite{4}-\cite{12}. Thus, there is still no clear answer to the
important problem of compactification and how contact can be made with a
realistic phenomenology.

In this paper, and as a toy model, we consider a closed bosonic string and
study the effect of successive toroidal compactifications.

In section $2$, we present general solutions resulted from various types of
an even dimensional tori compactifications. In section $3$, we display our
results and draw our conclusions.

\section{FORMALISM}

The Nambu-Goto action of a closed bosonic string is given by\cite{2},\cite{3}
:

\begin{equation}
S=-\frac 1{2\pi \alpha ^{\prime }}\int d\tau d\sigma [(x\cdot x^{\prime })^2-%
\stackrel{.}{x}^2\cdot x^{\prime 2}]
\end{equation}
with :

\begin{equation}
x^\mu (\sigma +\pi ,\tau )=x^\mu (\sigma ,\tau )
\end{equation}
and $\sigma ,\tau $ are the dimensionless world-sheet parameters. Here $%
\alpha ^{\prime }$ is the string scale and $x^{\prime \mu }$ (resp. $%
\stackrel{.}{x}^\mu $) means $\frac{\partial x^\mu }{\partial \sigma }$
(resp. $\frac{\partial x^\mu }{\partial \tau }$). The general solution of
the equation of motion (in the orthonormal gauge)

\begin{equation}
\stackrel{..}{x^\mu }-x"^\mu =0
\end{equation}
which satisfies the boundary condition (\ref{2}) is :

\begin{equation}
x^\mu (\sigma ,\tau )=q^\mu +\alpha ^{\prime }p^\mu +\frac i2\stackunder{%
n\neq 0}{\sum }\frac 1n[\alpha _n^\mu \exp -2in(\tau -\sigma )+\widetilde{%
\alpha }_n^\mu \exp -2in(\tau +\sigma )]
\end{equation}
where $q^\mu $ and $p^\mu $ are the string center of mass coordinates and
the momentum respectively.

After quantization, the critical dimension is fixed to $D=26$ and the
physical states $\left| \psi \right\rangle _{phys}$ are subject to the
Virasoro conditions :

\begin{eqnarray}
L_n\left| \psi \right\rangle _{phys} &=&\widetilde{L}_n\left| \psi
\right\rangle _{phys}=0\qquad \qquad n\geq 1 \\
(L_0-\widetilde{L}_0)\left| \psi \right\rangle _{phys} &=&0  \nonumber \\
(L_0+\widetilde{L}_0-\alpha (0))\left| \psi \right\rangle _{phys} &=&0 
\nonumber
\end{eqnarray}
where the Virasoro generators $L_n$ and $\widetilde{L}_n$ are given by : 
\begin{eqnarray}
L_n &=&\frac 1{4\alpha ^{\prime }}\stackrel{+\infty }{\stackunder{m=-\infty 
}{\sum }}\alpha _{n-m}\alpha _m \\
\widetilde{L}_n &=&\frac 1{4\alpha ^{\prime }}\stackrel{+\infty }{%
\stackunder{m=-\infty }{\sum }}\widetilde{\alpha }_{n-m}\widetilde{\alpha }_m
\nonumber
\end{eqnarray}
(here $\alpha (0)=2$). To get the mass spectrum, one has to apply the
following mass operator $M^2$

\[
M^2=4[N+\widetilde{N}-\alpha (0)] 
\]
on the physical states $\left| \psi \right\rangle _{phys}$ (we have taken $%
\frac 1{2\alpha ^{\prime }}=1$) with :

\[
N\left| \psi \right\rangle _{phys}=\widetilde{N}\left| \psi \right\rangle
_{phys} 
\]
where

\begin{eqnarray}
N &=&\stackrel{+\infty }{\stackunder{m=-\infty }{\sum }}\alpha _{-n}^\mu
\alpha _{n\mu } \\
\widetilde{N} &=&\stackrel{+\infty }{\stackunder{m=-\infty }{\sum }}%
\widetilde{\alpha }_{-n}^\mu \widetilde{\alpha }_{n\mu }  \nonumber
\end{eqnarray}

Now, our compactification program consists of starting from the critical
dimension $D=26$ and then truncating the extra dimensions successively
through a various number of tori compactifications.

We remained the reader that an $r-$dimensional torus $T^r$ is defined as the
set $R/\Gamma $, where $\Gamma $ is an $r-$dimensional lattice generated by
a basis $\{\overrightarrow{e_\alpha },\alpha =\overline{1,r}\}$. One can
also define a dual lattice $\Gamma ^{*}$ as

\begin{equation}
\Gamma ^{*}=\left\{ \overrightarrow{\beta }\in R^r/\forall \overrightarrow{%
\gamma }\in \Gamma ,\overrightarrow{\beta }\cdot \overrightarrow{\gamma }%
\qquad \text{is an integer }\right\}
\end{equation}
with a dual basis $\{\overrightarrow{e_\alpha ^{*}},\alpha =\overline{1,r}\}$
such that

\begin{equation}
\overrightarrow{e_\alpha ^{*}}\cdot \overrightarrow{e_\beta }=\delta
_{\alpha \beta }
\end{equation}

\subsection{METHOD $N=1$}

The first method consists of taking the left and right movers modes mixed.
Thus, the compactified coordinates $x$ can be written as :

\begin{equation}
x^i(\sigma ,\tau )=q^i+\alpha ^{\prime }p^i+\frac i2\stackunder{n\neq 0}{%
\sum }\frac 1n[\alpha _n^i\exp -2in(\tau -\sigma )+\widetilde{\alpha }%
_n^i\exp -2in(\tau +\sigma )]
\end{equation}
For the compactified coordinates $x^I$ $(I=\overline{1,r})$ on an $r-$%
dimensional torus, one has to identify the points under the translation by $%
2\pi R_\alpha $ in the $\overrightarrow{e_\alpha }$ direction. Thus :

\begin{equation}
x^I\simeq x^I+\frac \pi {\sqrt{2}}\stackunder{\alpha =1}{\stackrel{r}{\sum }}%
n_\alpha \cdot R_\alpha \cdot e_\alpha ^I\qquad (n\in Z)
\end{equation}
where $r$ (resp.$R_\alpha $ ) is the torus dimension (resp. radius in the $%
\alpha $ direction) and therefore one can write :

\begin{equation}
x^I(\sigma ,\tau )=q^I+\alpha ^{\prime }p^I+2\ell ^I++\frac i2\stackunder{%
n\neq 0}{\sum }\frac 1n[\alpha _n^I\exp -2in(\tau -\sigma )+\widetilde{%
\alpha }_n^I\exp -2in(\tau +\sigma )]
\end{equation}
with :

\begin{equation}
p^I=\stackunder{\alpha =1}{\stackrel{r}{\sum }}\frac{m_\alpha }{R_\alpha }%
\frac{e_\alpha ^{I*}}{\left\| e_\alpha ^{*}\right\| }\qquad \qquad (m_\alpha
\in Z)
\end{equation}
and $\ell $ are the winding numbers which have the following expression :

\begin{equation}
\ell ^I=\stackunder{\beta =1}{\stackrel{r}{\sum }}m_\beta R_\beta \frac{%
e_\beta ^I}{\left\| e_\beta \right\| }
\end{equation}

Now, after $"n"$ compactification, the mass operator $M$ takes the form :

\begin{equation}
M^2=4[N+\widetilde{N}-2\stackunder{k=1}{\stackrel{r}{\sum }}\stackunder{I=1}{%
\stackrel{r_k}{\sum }}(\frac{(p^I)^2}4+\ell ^{I^2})]
\end{equation}
(here $r_k$ is the dimension of the $k^{th}$ torus ($\stackunder{p=1}{%
\stackrel{n}{\sum }}r_p=22$ )). with :

\begin{equation}
N\left| \psi \right\rangle _{phys}=(\widetilde{N}+\stackunder{k=1}{\stackrel{%
n}{\sum }}\stackunder{I=1}{\stackrel{r_k}{\sum }}\ell ^Ip^I)\left| \psi
\right\rangle _{phys}
\end{equation}
where

\begin{eqnarray}
N &=&\stackrel{+\infty }{\stackunder{m=1}{\sum }}\stackunder{k=1}{\stackrel{n%
}{\sum }}\stackunder{I=1}{\stackrel{r_k}{\sum }}\alpha _{-m}^i\alpha _m^i+%
\stackrel{+\infty }{\stackunder{m=1}{\sum }}\stackunder{k=1}{\stackrel{n}{%
\sum }}\stackunder{I=1}{\stackrel{r_k}{\sum }}\alpha _{-m}^I\alpha _m^I \\
\widetilde{N} &=&\stackrel{+\infty }{\stackunder{m=1}{\sum }}\stackunder{k=1%
}{\stackrel{n}{\sum }}\stackunder{I=1}{\stackrel{r_k}{\sum }}\widetilde{%
\alpha }_{-m}^i\widetilde{\alpha }_m^i+\stackrel{+\infty }{\stackunder{m=1}{%
\sum }}\stackunder{k=1}{\stackrel{n}{\sum }}\stackunder{I=1}{\stackrel{r_k}{%
\sum }}\widetilde{\alpha }_{-m}^I\widetilde{\alpha }_m^I  \nonumber
\end{eqnarray}
with a self dual lattice and orthonormal basis, eqs.(\ref{16}) and (\ref{17}%
) become :

\begin{eqnarray}
M^2 &=&(N+\widetilde{N}-2)+\stackunder{k=1}{\stackrel{n}{\sum }}\stackunder{%
I=1}{\stackrel{r_k}{\sum }}(\frac{m_\alpha ^2}{R_\alpha ^2}+4n_\alpha
^2R_\alpha ^2) \\
N\left| \psi \right\rangle _{phys} &=&(\widetilde{N}+\stackunder{k=1}{%
\stackrel{n}{\sum }}\stackunder{I=1}{\stackrel{r_k}{\sum }}n_\alpha m_\alpha
)\left| \psi \right\rangle _{phys}  \nonumber
\end{eqnarray}

It is to be noted that one can characterizes the quantum physical states $%
\left| \psi \right\rangle _{phys}$ by the quantum numbers $n_\alpha $ and $%
m_\alpha $. Now, it is easy to show, that for $R_\alpha =\frac 1{\sqrt{2}%
}(\forall \alpha =\overline{1,r_p};p=\overline{1,n})$, the number of the
vectorial physical massless states and the quantum numbers $n_\alpha $ and $%
m_\alpha $ is (see APPENDIX A):

\begin{equation}
\Omega =4\stackunder{p=1}{\stackrel{n}{\sum }}(r_p^2+11)
\end{equation}
and

\begin{equation}
\Sigma =2\stackunder{p=1}{\stackrel{n}{\sum }}r_p=44
\end{equation}
respectively. However, for at least one $R_\alpha \neq \frac 1{\sqrt{2}}$,
the number of the physical vectorial massless states becomes $44$.

\subsection{METHOD $N=2$:}

In this method the left and the right movers of the closed string $%
x^I(\sigma -\tau )$ and $x^I(\sigma +\tau )$ respectively are treated
independently. In this case, the compactified coordinates can be written as :

\begin{eqnarray*}
x^I(\sigma -\tau ) &=&q^I+p^I(\tau -\sigma )+\frac i2\stackunder{n\neq 0}{%
\sum }\frac 1n\alpha _n^I\exp -2in(\tau -\sigma ) \\
x^I(\sigma +\tau ) &=&\widetilde{q}^I+\widetilde{p}^I(\tau +\sigma )+\frac i2%
\stackunder{n\neq 0}{\sum }\frac 1n\widetilde{\alpha }_n^I\exp -2in(\tau
+\sigma )
\end{eqnarray*}
where the string center of mass momentum $p^I$ and $\widetilde{p}^I$ are
given on the dual lattice $\Gamma ^{*}$ by : 
\begin{equation}
\begin{array}{c}
p^I=\stackunder{\alpha =1}{\stackrel{r}{\sum }}\frac{m_\alpha }{R_\alpha }%
\frac{e_\alpha ^{I*}}{\left\| e_\alpha ^{*}\right\| }\qquad \\ 
\widetilde{p}^I=\stackunder{\alpha =1}{\stackrel{r}{\sum }}\frac{\widetilde{m%
}_\alpha }{R_\alpha }\frac{e_\alpha ^{I*}}{\left\| e_\alpha ^{*}\right\| }%
\qquad
\end{array}
(m_\alpha \widetilde{m}_\alpha \in Z)  \label{22}
\end{equation}
It is to be noted that in this case, the winding numbers $\ell ^I$ and $%
\widetilde{\ell }^I$ are related to the momenta $p^I$ and $\widetilde{p}^I$
by the relations : 
\begin{equation}
\begin{array}{l}
\ell ^I=-\frac 12p^I \\ 
\widetilde{\ell }^I=\frac 12\widetilde{p}^I
\end{array}
\label{23}
\end{equation}
this means that the lattice $\Gamma $ and its dual $\Gamma ^{*}$ have a non
zero intersection. Now, the mass shell condition $(2-5-c)$ leads to the
relation 
\begin{equation}
M^2=-p^{i2}=2(N+\widetilde{N}-4)+\stackunder{k=1}{\stackrel{n}{\sum }}%
\stackunder{\beta ,\alpha =1}{\stackrel{r_k}{\sum }}[\frac{g_{\alpha \beta
}^{*}}{R_\alpha R_\beta }(m_\alpha m_\beta +\widetilde{m}_\alpha \widetilde{m%
}_\beta )]
\end{equation}
where $g_{\alpha \beta }^{*}$ is the dual lattice metric. Moreover, the
Virasoro condition $(2-5-b)$ implies that : 
\begin{equation}
(N+\frac 14\stackunder{k=1}{\stackrel{n}{\sum }}\stackunder{\beta ,\alpha =1%
}{\stackrel{r_k}{\sum }}\frac{g_{\alpha \beta }^{*}}{R_\alpha R_\beta }%
m_\alpha m_\beta )\left| \psi \right\rangle _{phys}=(\widetilde{N}+\frac 14%
\stackunder{k=1}{\stackrel{n}{\sum }}\stackunder{\beta ,\alpha =1}{\stackrel{%
r_k}{\sum }}\frac{g_{\alpha \beta }^{*}}{R_\alpha R_\beta }\widetilde{m}%
_\alpha \widetilde{m}_\beta )\left| \psi \right\rangle _{phys}
\end{equation}
It is important to mention that if $R_\alpha =\frac 1{\sqrt{2}}$, the
massless vectorial states belong to the adjoint representation of the
tensorial product $G\otimes G$, where $G$, is the simply laced Lie group of
rank $r=22$ and with a Cartan matrix $g_{\alpha \beta }$. Now, if the
lattice $\Gamma $ is even and integer i.e.: 
\[
\begin{array}{c}
\forall \overrightarrow{\beta },\overrightarrow{\gamma }\in \Gamma
\longrightarrow \overrightarrow{\beta }\cdot \overrightarrow{\gamma }\qquad 
\text{is an integer.} \\ 
\forall \overrightarrow{\gamma }\in \Gamma \longrightarrow \overrightarrow{%
\gamma ^2}\qquad \text{is integer and even.}
\end{array}
\]
the momenta $p^I$ and $\widetilde{p}^I$ are identified with the weight
vectors of the Lie group $G$. Now, if we characterize the vectorial physical
states by the quantum numbers $m_\alpha $ and $\widetilde{m}_\alpha $, we
can show that for $R_\alpha ^2=R^2=$ an integer or half integer the number $%
\Omega $ of these independent states is [see APPENDIX B] 
\begin{equation}
\Omega _2=44+\stackunder{p=1}{\stackrel{n}{\sum }}\frac{2^{s_{p+1}}}{%
(r_{p-}S_p)!Q_1!Q_2!....Q_{t_p}!}  \label{26}
\end{equation}
($n$ is the number of successive compactifications) where for the $p^{th}$
compactification ; $r_p$ ,$S_p$ and $Q_{t_p}$ are the dimension of the
compactified space, the number of the non zero quantum numbers ($m_\alpha $
and $n_\alpha $) and the degeneracy of the $t_p^{th}$ quantum number
respectively. However, if at least one of the $R$ is not an integer or half
an integer, the number, of the physical states becomes $\Omega _2^{\prime
}=44$.

\section{RESULTS AND CONCLUSIONS :}

To get an idea and keep our results transparent, we have considered
compactifications on an even dimensional tori. Tables $1$ and $2$ display
various types of compactifications and the rank and order of the resulted
Lie groups with both methods $1$ and $2$ with $R=\frac 1{\sqrt{2}}$and $1$
respectively. It is important to notice that the results depend on :

a) The choice of the method :

In fact, it is clear from tables $1$ and $2$ that for the same type of
compactification, the resulted Lie groups obtained with the first method are
totally different from the second one. For example, a compactification on $%
T^{22}$ gives with the first method the following possible Lie groups : $%
SO(5)\otimes SO(60)\otimes U(12)$; $SO(58)\otimes SO(22)\otimes U(4)$; $%
SO(14)\otimes SO(14)\otimes SO(61)$; $SO(56)\otimes SO(5)\otimes SO(29)$; $%
SO(63)\otimes U(8)\otimes U(5)$; $SO(44)\otimes SO(44)$; $SO(58)\otimes
SO(16)\otimes E7$. However, with the second method one gets : $SO(51)\otimes
SO(36)\otimes U(1)$; $SO(36)\otimes SO(3)\otimes SO(51);$ $SO(45)\otimes
SO(45)$. As a second example, the ten successive tori compactifications $%
T^2\otimes T^2\otimes T^2\otimes T^2\otimes T^2\otimes T^2\otimes T^2\otimes
T^2\otimes T^2\otimes T^4$ lead to no solutions for the first method and $%
SO(11)\otimes SO(58)\otimes U(12);$ $SO(47)\otimes SO(40)\otimes U(1);$ $%
SO(57)\otimes SO(24)\otimes U(4)$; $SO(59)\otimes SU(3)\otimes SO(14)$; $%
SO(3)\otimes SO(37)\otimes SO(51)$; $SO(58)\otimes SO(11)\otimes U(5)$; $%
SO(48)\otimes SO(36)\otimes G2$; $SO(58)\otimes SU(5)\otimes U(11)$ for the
second one.

b)Type and number of compactifications :

Each type and number of successive compactifications gives different
results. In fact, the type $T^2\otimes T^{20}$ (for example) leads to the
following Lie groups : $SO(12)\otimes SO(28)\otimes SO(48)$; $SO(38)\otimes
SO(42)\otimes U(4)$; $SO(8)\otimes SO(33)\otimes SO(49)$; $SO(55)\otimes
U(2)\otimes U(15)$; $SO(48)\otimes SO(28)\otimes E6$; $SO(48)\otimes
SO(26)\otimes E7$; However, the type $T^2\otimes T^2\otimes T^6\otimes
T^{12} $ gives $SO(20)\otimes U(15)\otimes U(19);$ $SO(13)\otimes
U(17)\otimes U(21) $; $SO(25)\otimes U(14)\otimes U(18)$.

c)Tori radius :

The results of the successive compactifications depend strongly on the
choice of the radius of the compactified tori. For example the first method
gives for $R=\frac 1{\sqrt{2}}(\forall \alpha =\overline{1,r};k=\overline{1,n%
})$, a number of $44+2\stackrel{n}{\stackunder{k=1}{\sum }}r_k^2$ vectorial
physical states which can form the irreducible representation of a Lie
group. However, for at least $R\neq \frac 1{\sqrt{2}}$, this number is
reduced to $44$ and leads to different Lie groups solutions.

\section{ACKNOWLEDGMENTS:}

We are very grateful to Drs. M. Lagraa and M. Tahiri for fruitful
discussions . \newpage\

\appendix\ 

The possible physical vectorial states are : $\alpha _{-t}^i\left|
0\right\rangle $ ,$\widetilde{\alpha }_{-t}^i\left| 0\right\rangle $,$\alpha
_{-t}^i\stackrel{q}{\stackunder{s=1}{\prod }}\widetilde{\alpha }%
_{-u_s}^{I_s}\left| 0\right\rangle $, $\widetilde{\alpha }_{-t}^i\stackrel{q%
}{\stackunder{s=1}{\prod }}\alpha _{-u_s}^{I_s}\left| 0\right\rangle $, $%
\widetilde{\alpha }_{-t}^i\stackrel{q}{\stackunder{s=1}{\prod }}\widetilde{%
\alpha }_{-u_s}^{I_s}\left| 0\right\rangle $ and $\alpha _{-t}^i\stackrel{q}{%
\stackunder{s=1}{\prod }}\alpha _{-u_s}^{I_s}\left| 0\right\rangle $ with $t$%
, $u$ and $q\in N^{*}$. For the states $\left| \psi \right\rangle _{phys}$
of the form $\alpha _{-t}^i\left| 0\right\rangle $ and $\widetilde{\alpha }%
_{-t}^i\left| 0\right\rangle $ and by imposing 
\begin{equation}
M^2\left| \psi \right\rangle _{phys}=0  \tag{A-1}
\end{equation}
and 
\begin{equation}
N\left| \psi \right\rangle _{phys}=(\widetilde{N}+\stackunder{p=1}{\stackrel{%
n}{\sum }}\stackunder{\alpha =1}{\sum }n_\alpha m_\alpha )\left| \psi
\right\rangle _{phys}  \tag{A-2}
\end{equation}
one gets 
\begin{equation}
\stackunder{p=1}{\stackrel{n}{\sum }}\stackunder{\alpha =1}{\stackrel{r_p}{%
\sum }}(\frac{m_\alpha ^2}{4R_\alpha ^2}+n_\alpha ^2R_\alpha ^2)=2-t 
\tag{A-3}
\end{equation}
and 
\[
\stackunder{p=1}{\stackrel{n}{\sum }}\stackunder{\alpha =1}{\stackrel{r_p}{%
\sum }}n_\alpha m_\alpha =\pm t 
\]
where $n$ denotes the number of tori compactifications ( $+$ and $-$ signs
are for the case $\alpha _{-t}^i\left| 0\right\rangle $ and $\widetilde{%
\alpha }_{-t}^i\left| 0\right\rangle $respectively). This implies that $t=1$
and $R=(2)^{-1/2}$ and one of the $n_\alpha $ and $m_\alpha $ are equal to $%
\pm 1$ (For the others, $n_\alpha =m_\beta $ if $\alpha \#\beta $). Thus,
the $4\stackunder{p=1}{\stackrel{n}{\sum }}r_p^2$ states can be written as:

$\left| 1,0,...,0;1,0,...,0\right\rangle $,$\left|
0,1,0,...,0;0,1,0,...,0\right\rangle ,...,$

$\left| -1,0,...,0;-1,0,...,0\right\rangle $,$\left|
0,-1,0,...,0;0,-1,0....,0\right\rangle $

$\left| 1,0,...,0;-1,0,...,0\right\rangle $,$\left|
0,1,0,...,0;0,-1,0,...,0\right\rangle ,...,$ \\and

$\left| -1,0,...,0;1,0,...,0\right\rangle $,$\left|
0,-1,0,...,0;0,1,0,...,0\right\rangle $,...,

It is worth to mention that the states of the form: 
\[
\widetilde{\alpha }_{-t}^i\stackrel{q}{\stackunder{s=1}{\prod }}\widetilde{%
\alpha }_{-u_s}^{I_s}\left| 0\right\rangle \text{ and }\alpha _{-t}^i%
\stackrel{q}{\stackunder{s=1}{\prod }}\alpha _{-u_s}^{I_s}\left|
0\right\rangle 
\]
verifying eq.$(A-1)$ and $\left( A-2\right) $ can be easily shown to be
equivalent to the states $\alpha _{-t}^i\left| 0\right\rangle $ and $%
\widetilde{\alpha }_{-t}^i\left| 0\right\rangle $ respectively. For the
states of the form $\alpha _{-t}^i\stackrel{q}{\stackunder{s=1}{\prod }}%
\widetilde{\alpha }_{-u_s}^{I_s}\left| 0\right\rangle $, $\widetilde{\alpha }%
_{-t}^i\stackrel{q}{\stackunder{s=1}{\prod }}\alpha _{-u_s}^{I_s}\left|
0\right\rangle $, the conditions $(A-2)$ and $(A-3)$ lead to : 
\[
2-(t+u_1+u_2+...+u_q)=\stackunder{p=1}{\stackrel{n}{\sum }}\stackunder{%
\alpha =1}{\stackrel{r_p}{\sum }}(\frac{m_\alpha ^2}{4R_\alpha ^2}+n_\alpha
^2R_\alpha ^2) 
\]
and 
\[
\stackunder{p=1}{\stackrel{n}{\sum }}\stackunder{\alpha =1}{\stackrel{r_p}{%
\sum }}n_\alpha m_\alpha =t-(u_1+u_2+...+u_q) 
\]
which implies that $n_\alpha =m_\alpha =0$ $(\alpha =\overline{1,r})$.
Thus,the number of the physical states is $2\stackunder{p=1}{\stackrel{n}{%
\sum }}r_p=44$. \newpage\ 

\begin{center}
\appendix\ 
\end{center}

The number of massless vectorial states of the form $\alpha _{-t}^i\left|
0\right\rangle $ , $\widetilde{\alpha }_{-t}^i\left| 0\right\rangle $ with $%
t\in N$ can be determined by solving the equations 
\begin{equation}
\begin{array}{c}
m_\alpha =0\qquad \qquad \qquad \stackunder{p=1}{\stackrel{n}{\sum }}%
\stackunder{\alpha =1}{\stackrel{r_p}{\sum }}\frac{m_\alpha ^2}{R_\alpha ^2}%
=2 \\ 
\widetilde{m}_\alpha =0\qquad \qquad \qquad \stackunder{p=1}{\stackrel{n}{%
\sum }}\stackunder{\alpha =1}{\stackrel{n_p}{\sum }}\frac{\widetilde{m}%
_\alpha ^2}{R_\alpha ^2}=2
\end{array}
\tag{B-1}
\end{equation}
respectively .Notice that in both cases the solution is the same .Setting $%
R_\alpha =R$ $(\forall \alpha =\overline{1,r};p=\overline{1,n})$ we obtain: 
\begin{equation}
\stackunder{p=1}{\stackrel{n}{\sum }}\stackunder{\alpha =1}{\stackrel{r_p}{%
\sum }}m_\alpha ^2=2R^2=\stackunder{p=1}{\stackrel{n}{\sum }}\stackunder{%
\alpha =1}{\stackrel{r_p}{\sum }}\widetilde{m}_\alpha ^2  \tag{B-2}
\end{equation}
Now,it is obvious that if $R^2$ is not an integer or half an integer, eqs.$%
(B-1)$ and $(B-2)$ have no solutions. In what follows we denote by $s$ the
number of the non zero $\widetilde{m}_\alpha $ 's. As an example, for $%
R=(2)^{3/2}$, eq.$(B-3)$ becomes 
\[
\stackunder{p=1}{\stackrel{n}{\sum }}\stackunder{\alpha =1}{\stackrel{r_p}{%
\sum }}\widetilde{m}_\alpha ^2=16 
\]
which can be written as: \\a)$1+1+1+1+1+1+1+1+1+1+1+1+1+1+1+1=16$ \\b)$%
4+1+1+1+1+1+1+1+1+1+1+1+1=16$ \\c)$4+4+1+1+1+1+1+1+1+1=16$ \\d)$%
4+4+4+1+1+1+1=16$ \\e)$4+4+4+4=16$ \\f)$9+1+1+1+1+1+1+1=16$ \\g)$%
9+4+1+1+1=16 $ \\In this case, the values of s are respectively $%
16,13,10,7,4,8,5$.

Now, if $R=1$ (case of our interest) one gets: 
\[
\stackunder{p=1}{\stackrel{n}{\sum }}\stackunder{\alpha =1}{\stackrel{r_p}{%
\sum }}\widetilde{m}_\alpha ^2=2 
\]
which implies that $\left| \widetilde{m}_\alpha \right| =1$. So, the
degeneracy $s$ of $\widetilde{m}_\alpha $ is equal to $2$. Thus the number $%
\Omega $ of all possible physical states of the form $\left|
11,1,0,...,0\right\rangle $,$\left| 11,0,1,...,0\right\rangle ,$...etc is 
\begin{equation}
\Omega =\frac 12\stackunder{p=1}{\stackrel{n}{\sum }}r_p(r_p-1)=\stackunder{%
p=1}{\stackrel{n}{\sum }}\frac{r_p!}{2!(r_p-1)}  \tag{B-3}
\end{equation}
This result can be found in an equivalent way by taking $r$ number arranged
in two and without repetition. Thus,the number of the different physical
vectorial states $\Omega $ is: 
\begin{equation}
\Omega =\stackunder{p=1}{\stackrel{n}{\sum }}\frac{2!C_{r_p}^2}{2!}=%
\stackunder{p=1}{\stackrel{n}{\sum }}\frac{r_p!}{2!(r_p-1)!}  \tag{B-4}
\end{equation}
Now, taking into account the positive and negative values of $\widetilde{m}%
_\alpha $ amounts to multiplying the result by $2^2$. Hence, the total
number, of states $\Omega _{tot}$ is: 
\begin{equation}
\Omega _{tot}=\stackunder{p=1}{\stackrel{n}{\sum }}2^2\frac{r_p!}{2!(r_p-2)!}
\tag{B-5}
\end{equation}
Then,it is clear that for a given $s$, eq.$(B-6)$ can be generalized to 
\[
\Omega _{tot}=\stackunder{p=1}{\stackrel{n}{\sum }}2^{s_p}\frac{r_p!}{%
(r_p-s_p)!\stackunder{q=1}{\stackrel{t_p}{\prod }}Q_q!} 
\]
where $Q_q$ (resp.$s_p$ ) is the degeneracy of the $q^{th}$ quantum number
(resp. the number of the non zero quantum numbers $\widetilde{m}_\alpha $ ), 
$t_p$ is the number of the non identical quantum numbers among the $s$ ones
for the $p^{th}$ compactification. The factor $\frac{r_p!}{(r_p-s_p)!}$
represents the number of the rearrangements of $r_p$ by $s_p$ numbers. i.e. 
\[
A_{r_p}^{s_p}=s_p!C_{r_p}^{s_p} 
\]
However, if there are some identical non zero quantum numbers, one has to
divide by the factor $\stackunder{q=1}{\stackrel{t_p}{\prod }}Q_q!$. Notice
that the factor $2^{s_p}$, comes from the fact that $\widetilde{m}_\alpha $
can take both positive and negative values.\newpage 

\begin{center}
{\bf TABLE CAPTION}
\end{center}

{\bf TABLE 1.} display the rank and order of the Lie groups coming from
various types of tori compactifications with the use of the first method and 
$R=\frac 1{\sqrt{2}}$.

{\bf TABLE 2.} the same as Table 1 but with the use of the second method and 
$R=1$.\newpage 

{\bf Table1:}\vspace{0.5cm} \\
\begin{tabular}{|lll|}
\hline
\multicolumn{1}{|l|}{\bf Type of compactification} & \multicolumn{1}{l|}{\bf %
rank} & {\bf order} \\ \hline
\multicolumn{1}{|l|}{$T^{22}$} & \multicolumn{1}{l|}{44} & 234212 \\ \hline
\multicolumn{1}{|l|}{$T^2\otimes T^{20}$} & \multicolumn{1}{l|}{44} & 174882
\\ \hline
\multicolumn{1}{|l|}{$T^4\otimes T^{18}$} & \multicolumn{1}{l|}{44} & 223644
\\ \hline
\multicolumn{1}{|l|}{$T^6\otimes T^{16}$} & \multicolumn{1}{l|}{44} & 123868
\\ \hline
\multicolumn{1}{|l|}{$T^8\otimes T^{14}$} & \multicolumn{1}{l|}{44} & 42172
\\ \hline
\multicolumn{1}{|l|}{$T^{10}\otimes T^{12}$} & \multicolumn{1}{l|}{44} & 
29116 \\ \hline
\multicolumn{1}{|l|}{$T^2\otimes T^2\otimes T^{18}$} & \multicolumn{1}{l|}{44
} & 115046 \\ \hline
\multicolumn{1}{|l|}{$T^2\otimes T^4\otimes T^{16}$} & \multicolumn{1}{l|}{44
} & 79758 \\ \hline
\multicolumn{1}{|l|}{$T^2\otimes T^6\otimes T^{14}$} & \multicolumn{1}{l|}{44
} & 47806 \\ \hline
\multicolumn{1}{|l|}{$T^2\otimes T^8\otimes T^{12}$} & \multicolumn{1}{l|}{44
} & 29630 \\ \hline
\multicolumn{1}{|l|}{$T^2\otimes T^{10}\otimes T^{10}$} & 
\multicolumn{1}{l|}{44} & 23742 \\ \hline
\multicolumn{1}{|l|}{$T^4\otimes T^4\otimes T^{14}$} & \multicolumn{1}{l|}{44
} & 46998 \\ \hline
\multicolumn{1}{|l|}{$T^4\otimes T^6\otimes T^{12}$} & \multicolumn{1}{l|}{44
} & 23254 \\ \hline
\multicolumn{1}{|l|}{$T^4\otimes T^8\otimes T^{10}$} & \multicolumn{1}{l|}{44
} & 15254 \\ \hline
\multicolumn{1}{|l|}{$T^6\otimes T^6\otimes T^{10}$} & \multicolumn{1}{l|}{44
} & 15126 \\ \hline
\multicolumn{1}{|l|}{$T^6\otimes T^8\otimes T^8$} & \multicolumn{1}{l|}{44}
& 12758 \\ \hline
\multicolumn{1}{|l|}{$T^2\otimes T^4\otimes T^4\otimes T^{12}$} & 
\multicolumn{1}{l|}{44} & 29408 \\ \hline
\multicolumn{1}{|l|}{$T^2\otimes T^4\otimes T^6\otimes T^{10}$} & 
\multicolumn{1}{l|}{44} & 12512 \\ \hline
\multicolumn{1}{|l|}{$T^2\otimes T^4\otimes T^8\otimes T^8$} & 
\multicolumn{1}{l|}{44} & 12832 \\ \hline
\multicolumn{1}{|l|}{$T^2\otimes T^2\otimes T^2\otimes T^{16}$} & 
\multicolumn{1}{l|}{44} & 89110 \\ \hline
\multicolumn{1}{|l|}{$T^2\otimes T^2\otimes T^4\otimes T^{14}$} & 
\multicolumn{1}{l|}{44} & 54640 \\ \hline
\multicolumn{1}{|l|}{$T^2\otimes T^2\otimes T^6\otimes T^{12}$} & 
\multicolumn{1}{l|}{44} & 32112 \\ \hline
\multicolumn{1}{|l|}{$T^2\otimes T^2\otimes T^8\otimes T^{10}$} & 
\multicolumn{1}{l|}{44} & 21400 \\ \hline
\multicolumn{1}{|l|}{$T^2\otimes T^6\otimes T^6\otimes T^8$} & 
\multicolumn{1}{l|}{44} & 11698 \\ \hline
\multicolumn{1}{|l|}{$T^4\otimes T^4\otimes T^4\otimes T^{10}$} & 
\multicolumn{1}{l|}{44} & 16820 \\ \hline
\multicolumn{1}{|l|}{$T^4\otimes T^4\otimes T^6\otimes T^8$} & 
\multicolumn{1}{l|}{44} & 11124 \\ \hline
\multicolumn{1}{|l|}{$T^4\otimes T^6\otimes T^6\otimes T^6$} & 
\multicolumn{1}{l|}{44} & 7732 \\ \hline
\multicolumn{1}{|l|}{$T^2\otimes T^2\otimes T^2\otimes T^2\otimes T^{14}$} & 
\multicolumn{1}{l|}{44} & 64602 \\ \hline
\multicolumn{1}{|l|}{$T^2\otimes T^2\otimes T^2\otimes T^4\otimes T^{12}$} & 
\multicolumn{1}{l|}{44} & 38898 \\ \hline
\multicolumn{1}{|l|}{$T^2\otimes T^2\otimes T^2\otimes T^6\otimes T^{10}$} & 
\multicolumn{1}{l|}{44} & 23898 \\ \hline
\multicolumn{1}{|l|}{$T^2\otimes T^2\otimes T^2\otimes T^8\otimes T^8$} & 
\multicolumn{1}{l|}{44} & 19074 \\ \hline
\multicolumn{1}{|l|}{$T^2\otimes T^2\otimes T^4\otimes T^4\otimes T^{10}$} & 
\multicolumn{1}{l|}{44} & 21706 \\ \hline
\multicolumn{1}{|l|}{$T^2\otimes T^2\otimes T^4\otimes T^6\otimes T^8$} & 
\multicolumn{1}{l|}{44} & 13970 \\ \hline
\multicolumn{1}{|l|}{$T^2\otimes T^2\otimes T^6\otimes T^6\otimes T^6$} & 
\multicolumn{1}{l|}{44} & 11066 \\ \hline
\multicolumn{1}{|l|}{$T^2\otimes T^4\otimes T^4\otimes T^4\otimes T^8$} & 
\multicolumn{1}{l|}{44} & 12530 \\ \hline
\multicolumn{1}{|l|}{$T^2\otimes T^4\otimes T^4\otimes T^6\otimes T^6$} & 
\multicolumn{1}{l|}{44} & 9674 \\ \hline
\multicolumn{1}{|l|}{$T^4\otimes T^4\otimes T^4\otimes T^4\otimes T^6$} & 
\multicolumn{1}{l|}{44} & 8250 \\ \hline
\end{tabular}
\newpage
\begin{tabular}{|lll|}
\hline
\multicolumn{1}{|l|}{$T^2\otimes T^2\otimes T^2\otimes T^2\otimes T^2\otimes
T^{12}$} & \multicolumn{1}{l|}{44} & 45424 \\ \hline
\multicolumn{1}{|l|}{$T^2\otimes T^2\otimes T^2\otimes T^2\otimes T^4\otimes
T^{10}$} & \multicolumn{1}{l|}{44} & 26866 \\ \hline
\multicolumn{1}{|l|}{$T^2\otimes T^2\otimes T^2\otimes T^2\otimes T^6\otimes
T^8$} & \multicolumn{1}{l|}{44} & 18034 \\ \hline
\multicolumn{1}{|l|}{$T^2\otimes T^2\otimes T^2\otimes T^4\otimes T^4\otimes
T^8$} & \multicolumn{1}{l|}{44} & 24612 \\ \hline
\multicolumn{1}{|l|}{$T^2\otimes T^2\otimes T^2\otimes T^4\otimes T^6\otimes
T^6$} & \multicolumn{1}{l|}{44} & 12916 \\ \hline
\multicolumn{1}{|l|}{$T^2\otimes T^2\otimes T^4\otimes T^4\otimes T^4\otimes
T^6$} & \multicolumn{1}{l|}{44} & 15990 \\ \hline
\multicolumn{1}{|l|}{$T^2\otimes T^4\otimes T^4\otimes T^4\otimes T^4\otimes
T^4$} & \multicolumn{1}{l|}{44} & 9464 \\ \hline
\multicolumn{1}{|l|}{$T^2\otimes T^2\otimes T^2\otimes T^2\otimes T^2\otimes
T^2\otimes T^{10}$} & \multicolumn{1}{l|}{44} & 30662 \\ \hline
\multicolumn{1}{|l|}{$T^2\otimes T^2\otimes T^2\otimes T^2\otimes T^2\otimes
T^4\otimes T^8$} & \multicolumn{1}{l|}{44} & 18590 \\ \hline
\multicolumn{1}{|l|}{$T^2\otimes T^2\otimes T^2\otimes T^2\otimes T^4\otimes
T^4\otimes T^6$} & \multicolumn{1}{l|}{44} & 12598 \\ \hline
\multicolumn{1}{|l|}{$T^2\otimes T^2\otimes T^2\otimes T^4\otimes T^4\otimes
T^4\otimes T^4$} & \multicolumn{1}{l|}{44} & 10510 \\ \hline
\multicolumn{1}{|l|}{$T^2\otimes T^2\otimes T^2\otimes T^2\otimes T^2\otimes
T^2\otimes T^2\otimes T^8$} & \multicolumn{1}{l|}{44} & 25910 \\ \hline
\multicolumn{1}{|l|}{$T^2\otimes T^2\otimes T^2\otimes T^2\otimes T^2\otimes
T^2\otimes T^4\otimes T^6$} & \multicolumn{1}{l|}{44} & 19616 \\ \hline
\multicolumn{1}{|l|}{$T^2\otimes T^2\otimes T^2\otimes T^2\otimes T^2\otimes
T^4\otimes T^4\otimes T^4$} & \multicolumn{1}{l|}{44} & 16458 \\ \hline
\multicolumn{1}{|l|}{$T^2\otimes T^2\otimes T^2\otimes T^2\otimes T^2\otimes
T^2\otimes T^2\otimes T^2\otimes T^6$} & \multicolumn{1}{l|}{44} & 23976 \\ 
\hline
\multicolumn{1}{|l|}{$T^2\otimes T^2\otimes T^2\otimes T^2\otimes T^2\otimes
T^2\otimes T^2\otimes T^4\otimes T^4$} & \multicolumn{1}{l|}{44} & 21848 \\ 
\hline
\multicolumn{1}{|l|}{$T^2\otimes T^2\otimes T^2\otimes T^2\otimes T^2\otimes
T^2\otimes T^2\otimes T^2\otimes T^2\otimes T^4$} & \multicolumn{1}{l|}{44}
& 28530 \\ \hline
\multicolumn{1}{|l|}{$T^2\otimes T^2\otimes T^2\otimes T^2\otimes T^2\otimes
T^2\otimes T^2\otimes T^2\otimes T^2\otimes T^2\otimes T^2$} & 
\multicolumn{1}{l|}{44} & 36180 \\ \hline
\end{tabular}
\newpage 

{\bf Table2:}\vspace{.5cm} \\
\begin{tabular}{|lll|}
\hline
\multicolumn{1}{|l|}{\bf Type of compactification} & \multicolumn{1}{l|}{\bf %
rank} & {\bf order} \\ \hline
\multicolumn{1}{|l|}{$T^{22}$} & \multicolumn{1}{l|}{44} & 9624428 \\ \hline
\multicolumn{1}{|l|}{$T^2\otimes T^{20}$} & \multicolumn{1}{l|}{44} & 6168104
\\ \hline
\multicolumn{1}{|l|}{$T^4\otimes T^{18}$} & \multicolumn{1}{l|}{44} & 6154058
\\ \hline
\multicolumn{1}{|l|}{$T^6\otimes T^{16}$} & \multicolumn{1}{l|}{44} & 1405082
\\ \hline
\multicolumn{1}{|l|}{$T^8\otimes T^{14}$} & \multicolumn{1}{l|}{44} & 568380
\\ \hline
\multicolumn{1}{|l|}{$T^{10}\otimes T^{12}$} & \multicolumn{1}{l|}{44} & 
243836 \\ \hline
\multicolumn{1}{|l|}{$T^2\otimes T^2\otimes T^{18}$} & \multicolumn{1}{l|}{44
} & 3841396 \\ \hline
\multicolumn{1}{|l|}{$T^2\otimes T^4\otimes T^{16}$} & \multicolumn{1}{l|}{44
} & 1821254 \\ \hline
\multicolumn{1}{|l|}{$T^2\otimes T^6\otimes T^{14}$} & \multicolumn{1}{l|}{44
} & 656222 \\ \hline
\multicolumn{1}{|l|}{$T^2\otimes T^8\otimes T^{12}$} & \multicolumn{1}{l|}{44
} & 317942 \\ \hline
\multicolumn{1}{|l|}{$T^2\otimes T^{10}\otimes T^{10}$} & 
\multicolumn{1}{l|}{44} & 189190 \\ \hline
\multicolumn{1}{|l|}{$T^4\otimes T^4\otimes T^{14}$} & \multicolumn{1}{l|}{44
} & 782618 \\ \hline
\multicolumn{1}{|l|}{$T^4\otimes T^6\otimes T^{12}$} & \multicolumn{1}{l|}{44
} & 305290 \\ \hline
\multicolumn{1}{|l|}{$T^4\otimes T^8\otimes T^{10}$} & \multicolumn{1}{l|}{44
} & 122554 \\ \hline
\multicolumn{1}{|l|}{$T^6\otimes T^6\otimes T^{10}$} & \multicolumn{1}{l|}{44
} & 111282 \\ \hline
\multicolumn{1}{|l|}{$T^6\otimes T^8\otimes T^8$} & \multicolumn{1}{l|}{44}
& 127674 \\ \hline
\multicolumn{1}{|l|}{$T^2\otimes T^4\otimes T^4\otimes T^{12}$} & 
\multicolumn{1}{l|}{44} & 425734 \\ \hline
\multicolumn{1}{|l|}{$T^2\otimes T^4\otimes T^6\otimes T^{10}$} & 
\multicolumn{1}{l|}{44} & 161392 \\ \hline
\multicolumn{1}{|l|}{$T^2\otimes T^4\otimes T^8\otimes T^8$} & 
\multicolumn{1}{l|}{44} & 91560 \\ \hline
\multicolumn{1}{|l|}{$T^2\otimes T^2\otimes T^2\otimes T^{16}$} & 
\multicolumn{1}{l|}{44} & 1283546 \\ \hline
\multicolumn{1}{|l|}{$T^2\otimes T^2\otimes T^4\otimes T^{14}$} & 
\multicolumn{1}{l|}{44} & 1041106 \\ \hline
\multicolumn{1}{|l|}{$T^2\otimes T^2\otimes T^6\otimes T^{12}$} & 
\multicolumn{1}{l|}{44} & 426592 \\ \hline
\multicolumn{1}{|l|}{$T^2\otimes T^2\otimes T^8\otimes T^{10}$} & 
\multicolumn{1}{l|}{44} & 186860 \\ \hline
\multicolumn{1}{|l|}{$T^2\otimes T^6\otimes T^6\otimes T^8$} & 
\multicolumn{1}{l|}{44} & 66458 \\ \hline
\multicolumn{1}{|l|}{$T^4\otimes T^4\otimes T^4\otimes T^{10}$} & 
\multicolumn{1}{l|}{44} & 156310 \\ \hline
\multicolumn{1}{|l|}{$T^4\otimes T^4\otimes T^6\otimes T^8$} & 
\multicolumn{1}{l|}{44} & 60072 \\ \hline
\multicolumn{1}{|l|}{$T^4\otimes T^6\otimes T^6\otimes T^6$} & 
\multicolumn{1}{l|}{44} & 37170 \\ \hline
\multicolumn{1}{|l|}{$T^2\otimes T^2\otimes T^2\otimes T^2\otimes T^{14}$} & 
\multicolumn{1}{l|}{44} & 1356836 \\ \hline
\multicolumn{1}{|l|}{$T^2\otimes T^2\otimes T^2\otimes T^4\otimes T^{12}$} & 
\multicolumn{1}{l|}{44} & 572866 \\ \hline
\multicolumn{1}{|l|}{$T^2\otimes T^2\otimes T^2\otimes T^6\otimes T^{10}$} & 
\multicolumn{1}{l|}{44} & 228514 \\ \hline
\multicolumn{1}{|l|}{$T^2\otimes T^2\otimes T^2\otimes T^8\otimes T^8$} & 
\multicolumn{1}{l|}{44} & 134170 \\ \hline
\multicolumn{1}{|l|}{$T^2\otimes T^2\otimes T^4\otimes T^4\otimes T^{10}$} & 
\multicolumn{1}{l|}{44} & 224942 \\ \hline
\multicolumn{1}{|l|}{$T^2\otimes T^2\otimes T^4\otimes T^6\otimes T^8$} & 
\multicolumn{1}{l|}{44} & 93598 \\ \hline
\multicolumn{1}{|l|}{$T^2\otimes T^2\otimes T^6\otimes T^6\otimes T^6$} & 
\multicolumn{1}{l|}{44} & 51502 \\ \hline
\multicolumn{1}{|l|}{$T^2\otimes T^4\otimes T^4\otimes T^4\otimes T^8$} & 
\multicolumn{1}{l|}{44} & 84654 \\ \hline
\end{tabular}
\newpage
\begin{tabular}{|lll|}
\hline
\multicolumn{1}{|l|}{$T^2\otimes T^4\otimes T^4\otimes T^6\otimes T^6$} & 
\multicolumn{1}{l|}{44} & 48470 \\ \hline
\multicolumn{1}{|l|}{$T^4\otimes T^4\otimes T^4\otimes T^4\otimes T^6$} & 
\multicolumn{1}{l|}{44} & 38934 \\ \hline
\multicolumn{1}{|l|}{$T^2\otimes T^2\otimes T^2\otimes T^2\otimes T^2\otimes
T^{12}$} & \multicolumn{1}{l|}{44} & 790064 \\ \hline
\multicolumn{1}{|l|}{$T^2\otimes T^2\otimes T^2\otimes T^2\otimes T^4\otimes
T^{10}$} & \multicolumn{1}{l|}{44} & 313838 \\ \hline
\multicolumn{1}{|l|}{$T^2\otimes T^2\otimes T^2\otimes T^2\otimes T^6\otimes
T^8$} & \multicolumn{1}{l|}{44} & 170800 \\ \hline
\multicolumn{1}{|l|}{$T^2\otimes T^2\otimes T^2\otimes T^4\otimes T^4\otimes
T^8$} & \multicolumn{1}{l|}{44} & 153804 \\ \hline
\multicolumn{1}{|l|}{$T^2\otimes T^2\otimes T^2\otimes T^4\otimes T^6\otimes
T^6$} & \multicolumn{1}{l|}{44} & 76606 \\ \hline
\multicolumn{1}{|l|}{$T^2\otimes T^2\otimes T^4\otimes T^4\otimes T^4\otimes
T^6$} & \multicolumn{1}{l|}{44} & 62226 \\ \hline
\multicolumn{1}{|l|}{$T^2\otimes T^4\otimes T^4\otimes T^4\otimes T^4\otimes
T^4$} & \multicolumn{1}{l|}{44} & 49194 \\ \hline
\multicolumn{1}{|l|}{$T^2\otimes T^2\otimes T^2\otimes T^2\otimes T^2\otimes
T^2\otimes T^{10}$} & \multicolumn{1}{l|}{44} & 430974 \\ \hline
\multicolumn{1}{|l|}{$T^2\otimes T^2\otimes T^2\otimes T^2\otimes T^2\otimes
T^4\otimes T^8$} & \multicolumn{1}{l|}{44} & 178496 \\ \hline
\multicolumn{1}{|l|}{$T^2\otimes T^2\otimes T^2\otimes T^2\otimes T^4\otimes
T^4\otimes T^6$} & \multicolumn{1}{l|}{44} & 89848 \\ \hline
\multicolumn{1}{|l|}{$T^2\otimes T^2\otimes T^2\otimes T^4\otimes T^4\otimes
T^4\otimes T^4$} & \multicolumn{1}{l|}{44} & 68048 \\ \hline
\multicolumn{1}{|l|}{$T^2\otimes T^2\otimes T^2\otimes T^2\otimes T^2\otimes
T^2\otimes T^2\otimes T^8$} & \multicolumn{1}{l|}{44} & 253958 \\ \hline
\multicolumn{1}{|l|}{$T^2\otimes T^2\otimes T^2\otimes T^2\otimes T^2\otimes
T^2\otimes T^4\otimes T^6$} & \multicolumn{1}{l|}{44} & 130820 \\ \hline
\multicolumn{1}{|l|}{$T^2\otimes T^2\otimes T^2\otimes T^2\otimes T^2\otimes
T^4\otimes T^4\otimes T^4$} & \multicolumn{1}{l|}{44} & 97954 \\ \hline
\multicolumn{1}{|l|}{$T^2\otimes T^2\otimes T^2\otimes T^2\otimes T^2\otimes
T^2\otimes T^2\otimes T^2\otimes T^6$} & \multicolumn{1}{l|}{44} & 187536 \\ 
\hline
\multicolumn{1}{|l|}{$T^2\otimes T^2\otimes T^2\otimes T^2\otimes T^2\otimes
T^2\otimes T^2\otimes T^4\otimes T^4$} & \multicolumn{1}{l|}{44} & 141310 \\ 
\hline
\multicolumn{1}{|l|}{$T^2\otimes T^2\otimes T^2\otimes T^2\otimes T^2\otimes
T^2\otimes T^2\otimes T^2\otimes T^2\otimes T^4$} & \multicolumn{1}{l|}{44}
& 198350 \\ \hline
\multicolumn{1}{|l|}{$T^2\otimes T^2\otimes T^2\otimes T^2\otimes T^2\otimes
T^2\otimes T^2\otimes T^2\otimes T^2\otimes T^2\otimes T^2$} & 
\multicolumn{1}{l|}{44} & 277918 \\ \hline
\end{tabular}

\end{document}